[

# Astrophysical Parameters of the Open Cluster NGC 2509

Talar YONTAN [*1] 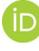, Seliz Koç[2]

**Abstract**

This study presents structural and fundamental astrophysical parameters of poorly studied open cluster NGC 2509. We used the third photometric and astrometric data release of the *Gaia* (*Gaia* DR3) to perform analyses. By taking into account the *Gaia* DR3 astrometric data, we calculated the membership probabilities of the stars in the region of NGC 2509. As a result of the membership analysis, 244 stars with membership probabilities $P \geq 50\%$ were determined as the physical members of the cluster. The colour excess, distance and age were obtained simultaneously by fitting solar metallicity PARSEC isochrones to $G \times G_{BP}$-$G_{RP}$ colour-magnitude diagram. We considered the most likely cluster member stars during the fitting procedure and calculated the colour excess, distance and age of the NGC 2509 as $E(G_{BP}$-$G_{RP}) = 0.100 \pm 0.015$ mag, $d = 2518 \pm 667$ pc and $t = 1.5 \pm 0.1$ Gyr, respectively.

## 1. INTRODUCTION

Open star clusters (OCs) are the celestial bodies which consist of tens of to thousands of stars that share similar physical properties under the weak gravitational forces. These objects are located through the Galactic disc within the different distances and have a vast age range from a few million years to a few billion years. Because of the components stars of OCs are formed from collapsing of the same molecular cloud, their age, heliocentric distance, metallicities are similar [1]. These properties make OCs important tools to study star formation process, stellar evolution as well as the formation and chemical evolution of the Galactic disc ([2-4]). Due to their same formation origin, the movement vectors of proper-motion components of cluster stars are similar. This knowledge is useful to define the membership properties of stars in the direction of open cluster ([5]). Moreover, astrometric, photometric and spectroscopic data of *Gaia* observations provided essential results for OCs and Milky Way Galaxy (e.g., [6-10]).

* Corresponding author: talar.yontan@istanbul.edu.tr (T. YONTAN)
[1] Istanbul University, Faculty of Science, Department of Astronomy and Space Sciences, Istanbul, Turkey.
[2] Istanbul University, Institute of Graduate Studies in Science, Programme of Astronomy and Space Sciences, Istanbul, Turkey.
E-mail: seliskoc@gmail.com
ORCID: https://orcid.org/ 0000-0002-5657-6194, https://orcid.org/ 0000-0001-7420-0994

Table 1 Fundamental parameters for NGC 2509 estimated in this study and compiled from the literature. Columns denote the colour excess ($E(B-V)$), metallicity ([Fe/H]), distance moduli and distance ($\mu$, $d$), age ($t$), proper-motion components ($\mu_\alpha\cos\delta$, $\mu_\delta$) and trigonometric parallaxes ($\varpi$). Errors of the parameters are shown in parenthesis

| $E(B-V)$ (mag) | [Fe/H] (dex) | $\mu$ (mag) | $d$ (pc) | $t$ (Gyr) | $\mu_\alpha\cos\delta$ (mas/yr) | $\mu_\delta$ (mas/yr) | $\varpi$ (mas) | Reference |
|---|---|---|---|---|---|---|---|---|
| 0.15 | --- | --- | 912 (15) | 8 | --- | --- | --- | [11] |
| 0.1 | --- | 12.5 (0.10) | 2900 | 1.2 | --- | --- | --- | [12] |
| 0.104 | --- | 11.2 | 1711 | 1.6 | -3.82 | 4.24 | --- | [13] |
| 0.104 | --- | --- | 1711 | 1.6 | --- | --- | --- | [14] |
| --- | --- | --- | 2549 (695) | --- | -2.708 (0.076) | 0.764 (0.075) | 0.363 (0.039) | [6] |
| --- | --- | --- | --- | 2.3 (0.140) | -2.712 (0.160) | 0.771 (0.138) | 0.369 (0.032) | [15] |
| 0.1 | --- | 11.72 | --- | 1.8 | --- | --- | --- | [16] |
| --- | --- | --- | 2549 (695) | --- | -2.708 (0.076) | 0.764 (0.075) | 0.363 (0.039) | [17] |
| 0.074 | --- | 11.99 | 2495 | 1.5 | -2.708 (0.076) | 0.764 (0.075) | 0.363 (0.039) | [18] |
| 0.242 | --- | 12.36 | --- | 0.86 | --- | --- | --- | [19] |
| 0.105 (0.021) | 0.082 (0.137) | --- | 2411 (118) | 1.5 | -2.708 (0.098) | 0.766 (0.089) | 0.364 (0.041) | [20] |
| 0.071 (0.011) | 0.00 | 12.191 (0.509) | 2518 (667) | 1.5 (0.1) | -2.718 (0.002) | 0.803 (0.002) | 0.37 (0.03) | This study |

NGC 2509 ($\alpha_{2000.0}$ = 08$^h$00$^m$48$^s$.2, $\delta_{2000.0}$ = -19°03'22"; $l$ = 237°.8442, $b$ = 05°.8465) [17] is an intermediate age open cluster located in the second Galactic quadrant towards the Galactic anti-centre region. Due to the cluster is situated very close to the Galactic disc, its embedded in field star contamination. [13] found reddening, distance and age of the cluster as $E(B-V)$ = 0.104 mag, $d$ = 1711 pc, log $t$ = 9.2 yr, respectively. [18] used Gaia DR2 astrometric and photometric data and determined distance and age of NGC 2509 as $d$ = 2495 pc and log $t$ = 9.18 yr, respectively. They obtained extinction in the cluster direction as $A_V$ = 0.23 which corresponds colour excess to be $E(B-V)$ = 0.074 mag. Also, researchers calculated mean proper-motion components of the NGC 2509 as ($\mu_\alpha\cos\delta$, $\mu_\delta$) = (-2.708±0.076, 0.764±0.075) mas/yr. [19] found extinction, distance and age of the cluster as $A_V$ = 0.75, which corresponds $E(B-V)$ to be 0.242 mag, $\mu$ = 12.36 mag and $t$ = 860 Myr, respectively. [20] using Gaia DR2 astrometric and photometric data obtained the extinction, metallicity, distance and age of the cluster as $A_V$ = 0.23 mag, which corresponds $E(B-V)$ to be 0.105 mag, [Fe/H] = 0.082±0.137 dex, $d$ = 2411±118 pc and log $t$ = 9.18 yr, respectively. Moreover, in the study researchers determined mean proper-motion components as ($\mu_\alpha\cos\delta$, $\mu_\delta$) = (-2.708±0.098, 0.766±0.089) mas/yr. For detailed literature study results see Table 1.

In this study, we used Gaia's the third release (hereafter Gaia DR3, [21]) of astrometric and photometric data to analyse NGC 2509 open cluster. To remove field star contamination and perform precise selection of cluster members we used Gaia DR3 proper-motion components ($\mu_\alpha\cos\delta$, $\mu_\delta$), trigonometric parallaxes ($\varpi$) and their uncertainties for each star in cluster region and calculated their membership probabilities. We took into account the stars with the membership probabilities $P \geq 50\%$ as the most likely cluster members and used them to determine astrometric ($\mu_\alpha\cos\delta$, $\mu_\delta$, $\varpi$) and astrophysical (reddening, distance modulus and age) parameters of the NGC 2509.

### 1.1. Astrometric and Photometric Data

We used photometric and astrometric data of Gaia DR3 ([21]) to analyse NGC 2509 open cluster. We created the catalogue of NGC 2509 by considering 20 arcmin radius circular field from the cluster centre ($\alpha_{2000.0}$ = $01^h51^m12^s.7$, $\delta_{2000.0}$ = 61°03'40"; [17]). Thus, we found 20,391 stars within the magnitude range $7 < G \leq 22$ mag in cluster region. Generated catalogue contains positions ($\alpha$, $\delta$), photometric magnitude and colours ($G$, $G_{BP}$-$G_{RP}$), proper-motion components ($\mu_\alpha \cos\delta$, $\mu_\delta$) and trigonometric parallaxes ($\varpi$). Identification map of the open cluster NGC 2509 is shown in Figure 1. *Gaia* DR3 ([21]) provides high quality astrometric and photometric data more than 1.5 billion celestial objects. The uncertainties in *Gaia* DR3 are 0.01-0.02 mas for $G \leq 15$ mag, and reach about 1 mas at $G = 21$ mag. The uncertainties of trigonometric parallax are 0.02-0.07 mas for $G \leq 17$ mag, 0.5 mas for $G = 20$ mag and reach 1.3 mas for at $G = 21$ mag. For sources with $G \leq 17$ mag the proper motion uncertainties are 0.02-0.07 mas/yr reaching 0.5 mas mas/yr at $G = 20$, and 1.4 mas/yr at $G = 21$ mag. For sources within $G \leq 17$ mag the $G$-band photometric uncertainties are 0.3-1 mmag, increasing to 6 mmag at $G = 20$ mag, allowing such separations to be made more accurately ([7, 22]).

.

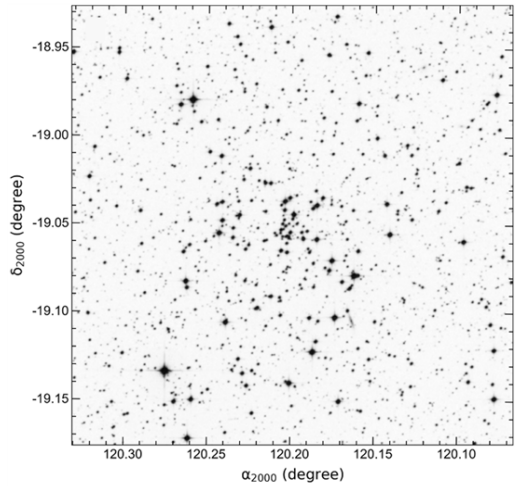

Figure 1 Identification chart of stars located through the area of NGC 2509. Field of view of the optical chart is 20' × 20'. North and East correspond to the up and left directions, respectively

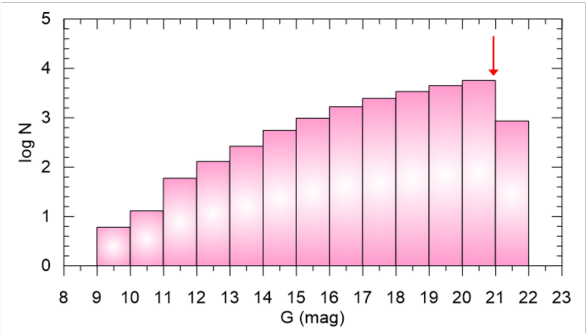

Figure 2 Number of stars versus interval $G$ magnitudes. The red arrow represents the faint limit magnitude of the NGC 2509

Table 2 Mean photometric errors for $G$ magnitude and $G_{BP}$-$G_{RP}$ colour index as $G$ mag function. $N$ indicates the number of stars within the selected $G$ magnitude intervals

| $G$ (mag) | $N$ | $\sigma_G$ (mag) | $\sigma_{GBP-GRP}$ (mag) |
|---|---|---|---|
| ( 6, 12] | 79 | 0.003 | 0.005 |
| (12, 14] | 395 | 0.003 | 0.005 |
| (14, 15] | 552 | 0.003 | 0.005 |
| (15, 16] | 967 | 0.003 | 0.006 |
| (16, 17] | 1643 | 0.003 | 0.008 |
| (17, 18] | 2474 | 0.003 | 0.016 |
| (18, 19] | 3418 | 0.003 | 0.034 |
| (19, 20] | 4456 | 0.004 | 0.074 |
| (20, 21] | 5563 | 0.009 | 0.188 |
| (21, 23] | 844 | 0.024 | 0.383 |

To perform precise analyses, we investigated photometric completeness limit for the cluster. For this, we determined number of stars in interval $G$ magnitudes. The star count histogram versus interval $G$ magnitude is shown in Figure 2. It can be seen from the figure that number of stars increase up to $G = 21$ mag and start to decrease after this limit where the stellar incompleteness has set in. This value is adopted photometric completeness limit for NGC 2509. We considered the stars fainter than this limit to utilize analyses of the cluster. We also investigated mean photometric errors of $G$ magnitudes and $G_{BP}$-$G_{RP}$ colour indices as function of $G$ interval magnitudes and listed in

Table 2. It can be seen from Table 2 that the mean errors of $G$ magnitude and $G_{BP}$-$G_{RP}$ colour indices of the stars reach up to 0.01 and 0.2 mag for completeness limit $G = 21$ mag, respectively

## 3 CONCLUSIONS AND DISCUSSION

### 3.1. Structural Parameters of the NGC 2509

In order to estimate structural parameters of NGC 2509, we visualized the Radial Density Profile (RDP) for NGC 2509 using the centre coordinates given by [17]. The cluster area was divided into concentric circles around the adopted centre. Stellar densities ($\rho_i$) in each $i^{th}$ ring were calculated by using the expression $\rho_i=N_i/A_i$, where $N$ is the number of stars and $A$ is the area of $i^{th}$ ring. Those calculated stellar densities ($\rho$) were plotted as a function of radius from the cluster centre (Figure 3). We fitted RDP model of [23] to this distribution considering $\chi^2$ minimization. The RDP model is formulated as $\rho(r) = f_{bg} + (f_0 / (1 + (r/r_c)^2)$, where $r$ represents the radius from the cluster centre, $f_{bg}$, $f_0$ and $r_c$ describe the background stellar density, the central density and the core radius, respectively. RDP of the cluster with best fit is shown in Figure 3. As a result of analyses, we estimated the core radius, background stellar density and central density of the NGC 2509 as $r_c = 0.578\pm0.113$ arcmin, $f_{bg}=15.816\pm0.511$ stars/arcmin$^2$ and $f_0 = 32.334\pm3.236$ stars/arcmin$^2$, respectively. Also, with visual review of RDP, we obtained limiting radii ($r_{lim}$) of the cluster. We described the $r_{lim}$ as the radius that cluster density almost meets the background density (Figure 3 grey horizontal line). We considered this limit to be $r_{lim} = 7'$ (5.13±1.35 pc) for the cluster, and we used the stars inside this limiting radius for the determination of fundamental parameters of it.

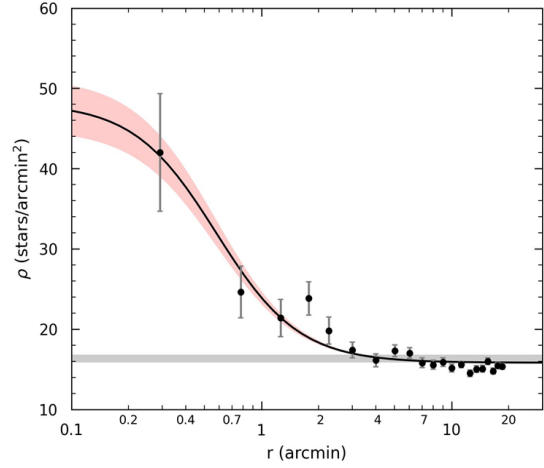

Figure 3 RDP of NGC 2509. Errors were calculated from sampling statistics $1/\sqrt{N}$, where $N$ is the number of stars used in the density estimation. The smooth line shows the best fit profile of [23]. The background density level and its errors are represented with the horizontal grey bands. The King fit uncertainty (1σ) is shown by the red shaded region

### 3.2. Membership Analyses

OCs are distributed through the densely populated Galactic plane and are mostly affected by foreground/background stars. It is necessary to separate cluster members from field stars to determine more precise fundamental parameters of the OCs. As for the cluster members have same origin, their motion vectors in the sky point in the same direction. Therefore, proper-motion components are useful tools to discriminate cluster members from field stars. The membership analyses carried from astrometric data of *Gaia* catalogue give more reliable results than ground-based data ([24]). We utilized UPMASK (Unsupervised Photometric Membership Assignment in Stellar Clusters; [29]) method by using astrometric data of *Gaia* DR3 for membership analyses. This methodology previously used in many studies ([6, 17, 25-27]). UPMASK is the method of clustering algorithm and detects spatially populated groups and identifies membership probabilities of stars. This clustering method is described as k-means clustering which is the integer number and varies within 5 to 25 and is not set directly by the user ([17, 28]). We used five-dimensional astrometric parameters as inputs which include stars' positions (α, δ), proper-motion components (μ$_α$cosδ, μ$_δ$), trigonometric parallaxes (ϖ) and their uncertainties. In the method of UPMASK, membership probabilities (P) of stars are described as the frequency with which star belongs to a clustered group. Program was run 1000 iterations during the utilization. Best solution was adopted when k-means value was set to 15 for NGC 2509. We considered the stars with membership probability $P \geq 50\%$ as possible cluster members. Moreover, to determine cluster members more precision and get reliable astrometric parameters for NGC 2509, we filtered possible members which brighter than completeness limit $G \leq 21$ mag and within the cluster limiting radius ($r_{lim}$=7'). Thus, with these criteria we identified a total of 244 stars with probability $P \geq 50\%$. We plotted proper-motion distribution of stars to image the position of the cluster as regards to field stars and shown it in Figure 4.

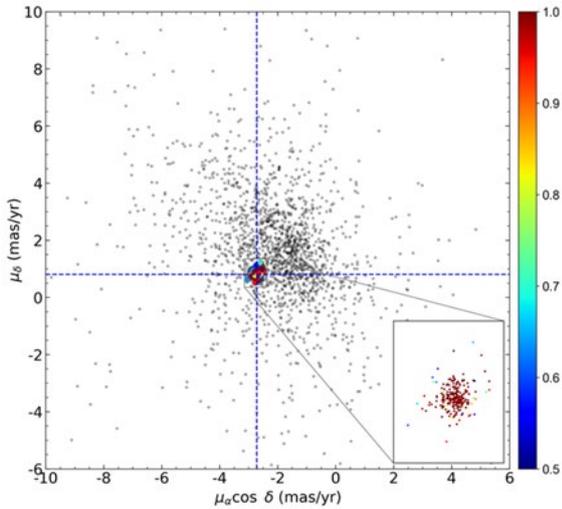

Figure 4 *Gaia* DR3 data based proper-motion distribution of NGC 2509. Colour bar shows the scale of membership probabilities of the stars. Zoomed region shows the location where the cluster is condensed. The intersection of the dashed blue lines is the point of mean proper-motion values

It can be seen in the figure that NGC 2509 is embedded in field stars. In Figure 4, the intersection of blue dashed lines represents the values of mean proper-motion components calculated from the most probable cluster members (244 stars with $P \geq 50\%$). These values were determined as (μ$_α$cosδ, μ$_δ$) = (-2.718±0.002, 0.803±0.002) mas/yr, which are compatible with the results of all studies performed with *Gaia* observations for the NGC 2509 (see Table 1). To estimate mean trigonometric parallax ⟨ϖ⟩ of NGC 2509, we took into account the most likely member stars with relative parallax error (σ$_ϖ$/ϖ) less than 0.15 and plotted the histogram for number of stars (N) versus trigonometric parallax (ϖ) as shown in Figure 5. We fitted Gaussian function to this distribution and it provided the mean trigonometric parallax ⟨ϖ⟩ to be 0.37±0.03 mas. By using the expression of linear equation $d$ (pc) = 1000/ϖ (mas) we calculated trigonometric parallax-based distance of the NGC 2509 as $d_ϖ$ = 2703±190 pc.

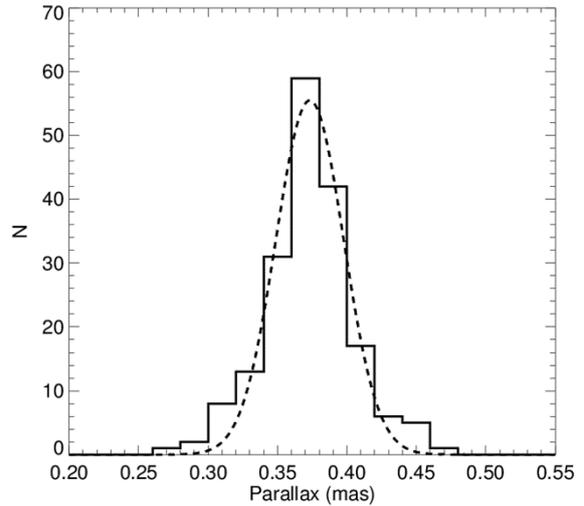

Figure 5 *Gaia* DR3 based trigonometric parallax histogram for NGC 2506. Applied Gaussian fit is shown in black dashed curve

### 3.3. Fundamental Parameters of NGC 2509

Colour-magnitude diagrams of OCs are important tools to examine morphology of the cluster, as well as to obtain their fundamental parameters such as reddening, metallicity, age and distance.

The reddening, metallicity age and distance of NGC 2509 were derived simultaneously by fitting PARSEC isochrones of [29] to the observed colour-magnitude diagram as shown in Figure 6. During the fitting procedure we concentrated on most likely stars that contain main-sequence, turn-off point and giant region of the cluster. Different age values log $t$ = 8.15, 8.18 and 8.20 yr with metallicity $z$ = 0.0152 of isochrones were superimposed on observable colour-magnitude diagram. Best fit isochrone resulted the reddening, age and isochrone distance of the NGC 2509 to be $E(G_{BP}-G_{RP})$ = 0.100±0.015 mag, $d_{iso}$ = 2518±667 pc and $t$ = 1.5±0.1 Gyr, respectively (see Table 1). To compare with literature, we transformed *Gaia* based reddening to *UBV* based value by using $E(G_{BP}-G_{RP})$ = 1.41× $E(B-V)$ equation given by [30]. This equation was used in the studies recently presented [31, 32] and resulted successfully. Thus, we calculated *UBV* based reddening as $E(B-V)$=0.071±0.011 mag. The $E(B-V)$ colour excess and age of the cluster derived in the study are in a good agreement with many values given in the literature (see Table 1). Moreover, the distance value ($d_{iso}$ = 2518±667 pc) found in study is compatible with the values those of represented in *Gaia* era (see Table 1), and with the distance that calculated from trigonometric parallaxes of most likely members in the study. This shows that the cluster's all astrophysical parameters derived in the study are acceptable. Also, results of the study listed in Table 3.

**4 CONCLUSION**

We performed *Gaia* DR3 astrometric and photometric data-based study of intermediate-age open cluster NGC 2509. We calculated membership probabilities of stars in the region of cluster and classified 244 most likely members with membership probabilities $P \geq 50\%$. We considered these members to estimate astrophysical parameters. The main results of the analyses are as follows:

- Taking into account the RDP results, the limiting radius of the cluster is obtained as $r_{lim}$=7′ (5.13±1.35 pc)

- Using the distribution of proper-motion components, we determined mean proper-motion of NGC 2509 as $\langle\mu_\alpha\cos\delta, \mu_\delta\rangle$ = (-2.718±0.002, 0.803±0.002) mas/yr.

- On the basis of most likely cluster stars with relative parallax error ($\sigma_\varpi/\varpi$) less than 0.15, we calculated trigonometric parallax of the cluster as $\langle\varpi\rangle$ = 0.37±0.03 mas, which corresponds the parallax distance to be $d_\varpi$ = 2703±190 pc.

- PARSEC isochrones of [30] provide an age of $t$ = 1.5±0.1 Gyr and isochrone distance of $d_{iso}$ = 2518±667 pc with the solar metallicity ($z$ = 0.015) for the cluster. Because of astrophysical parameters estimated simultaneously on CMD, the best fit isochrone gives the *Gaia* based reddening as $E(G_{BP}-G_{RP})$ = 0.100±0.015 mag for NGC 2509.

Table 3 Astrophysical parameters of NGC 2509 estimated in the study

| Parameter | Value |
| --- | --- |
| α (hh:mm:ss.s) | 08:00:48.2 |
| δ (dd:mm:ss.s) | -19:03:22.0 |
| $l$ (°) | 237.8442 |
| $b$ (°) | 05.8465 |
| $f_0$ (star/arcmin$^2$) | 32.334±3.236 |
| $r_c$ (arcmin) | 0.578±0.113 |
| $f_{bg}$ (star/arcmin$^2$) | 15.816±0.511 |
| $r_{lim}$ (arcmin) | 7 |
| $r$ (pc) | 5.13±1.35 |
| $\mu_\alpha\cos\delta$ (mas/yr) | -2.718±0.002 |
| $\mu_\delta$ (mas/yr) | 0.803±0.002 |
| Cluster members ($P \geq 50\%$) | 244 |
| $\varpi$ (mas) | 0.37±0.03 |
| $d_\varpi$ (pc) | 2703±190 |
| $E(G_{BP} - G_{RP})$ (mag) | 0.100±0.015 |
| $E(B-V)$ (mag) | 0.071±0.011 |
| [Fe/H] (dex) | 0.0152 |
| Age (Gyr) | 1.5±0.1 |
| Distance module (mag) | 12.191±0.509 |
| Isochrone distance (pc) | 2518±667 |
| $X$ (pc) | -1333 |
| $Y$ (pc) | -2121 |
| $Z$ (pc) | 256 |
| $R_{gc}$ (kpc) | 9.57 |

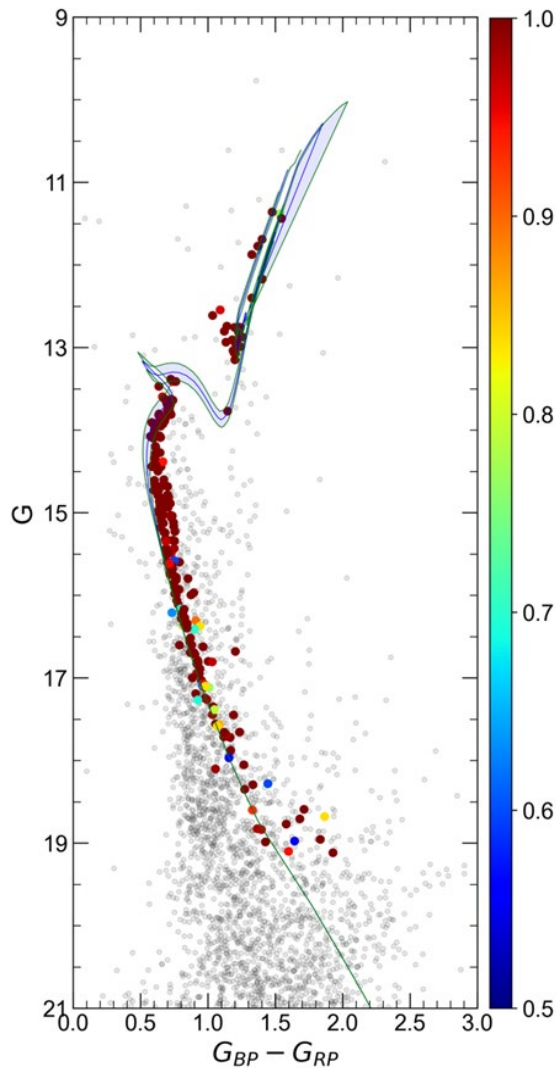

Figure 6 Colour-magnitude diagram with best fit PARSEC isochrones. Colour scaled represent the membership probabilities of stars according to colour bar in right panel. Gray circles show field stars. Blue and green solid lines represent errors in age, with isochrone that best represent cluster parameters, respectively

*Acknowledgments*
This study has been supported in part by the Scientific and Technological Research Council (TÜBİTAK) 122F109. We also made use of VizieR and Simbad databases at CDS, Strasbourg, France. We made use of data from the European Space Agency (ESA) mission Gaia, processed by the Gaia Data Processing and Analysis Consortium (DPAC). Funding for DPAC has been provided by national institutions, in particular the institutions participating in the Gaia Multilateral Agreement.

*Authors' Contribution*
Concept: T.Y., S.K., Design: T.Y., S.K., Data Collection or Processing: T.Y., S.K., Analysis or Interpretation T.Y., S.K., Literature Search: T.Y., S.K., Writing: T.Y., S.K.

*The Declaration of Conflict of Interest/ Common Interest*
The study is complied with research and publication ethics.

*The Declaration of Ethics Committee Approval*
This study does not require ethics committee permission or any special permission.

*The Declaration of Research and Publication Ethics*
The authors of the paper declare that they comply with the scientific, ethical and quotation rules of SAUJS in all processes of the paper and that they do not make any falsification on the data collected. In addition, they declare that Sakarya University Journal of Science and its editorial board have no responsibility for any ethical violations that may be encountered, and that this study has not been evaluated in any academic publication environment other than Sakarya University Journal of Science.

## REFERENCES

[1] Yadav, R. K. S., Glushkhova, E. V., Sariya, D. P., Porokhova, A. V., Kumar, B., Sagar, R., "Optical Photometric Study of the Open Clusters Koposov 12, Koposov 53 and Koposov 77", Monthly Notices of the Royal Astronomical Society, vol. 414, pp. 652-658, 2011.

[2] Lada, C. J., Lada, E. A., "Embedded Clusters in Molecular Clouds", Annual Review of Astronomy and Astrophysics, vol. 41, pp. 57-115, 2003.

[3] Kim, S. C., Kyeong, J., Park, H. S., Han, I., Lee, J. H., Moon, D., Lee, Y., Kim, S., "BVI Photometric Study of the Old

Open Cluster Ruprecht 6", Journal of the Korean Astronomical Society, vol. 50, pp. 79-92, 2017.

[4] He, Z.-H., Xu, Y., Hou, L.-G., "A Catalogue of 74 New Open Clusters Found in Gaia Data Release 2", Research in Astronomy and Astrophysics, vol. 21, article id.A93, pp. 1-9, 2021.

[5] Bisht, D., Elsanhoury, W. H., Zhu, Q., Sariya, D. P., Yadav, R. K. S., Rangwal, G., Durgapal, A., Jiang, I-G., "An Investigation of Poorly Studied Open Cluster NGC 4337 Using Multicolor Photometric and Gaia DR2 Astrometric Data", The Astronomical Journal, vol. 160, pp. 1-14, 2020.

[6] Cantat-Gaudin, T., Jordi, C., Vallenari, A., Bragaglia, A., Balaguer-Núñez, L., Soubiran, C., Bossini, D., Moitinho, A., Castro-Ginard, A., Krone-Martins, A., Casamiquela, L., Sordo, R., Carrera, R., "A Gaia DR2 View of the Open Cluster Population in the Milky Way", Astronomy and Astrophysics, vol. 618, article id.A93, pp. 1-16, 2018.

[7] Soubiran, C., Cantat-Gaudin, T., Romero-Gómez, M., Casamiquela, L., Jordi, C., Vallenari, A., Antoja, T., Balaguer-Núñez, L., Bossini, D., Bragaglia, A., Carrera, R., Castro-Ginard, A., Figueras, F., Heiter, U., Katz, D., Krone-Martins, A., Le Campion, J. -F., Moitinho, A., Sordo, R., "Open Cluster Kinematics with Gaia DR2", Astronomy and Astrophysics, vol. 619, article id.A155, pp. 1-11, 2018.

[8] Castro-Ginard, A., Jordi, C., Luri, X., Cantat-Gaudin, T., Balaguer-Núñez, L., "Hunting for Open Clusters in Gaia DR2: The Galactic Anticentre", Astronomy and Astrophysics, vol. 627, article id.A35, pp. 1-8, 2019.

[9] Ding, X., Ji, K-F., Li, X-Z., Cheng, Q-Y., Wang, J-L., Yu, X-G., Liu, H., "Fundamental Parameters for 30 Faint Open Clusters with Gaia EDR3 Based on the More Reliable Members", Publications of the Astronomical Society of Japan, vol. 73, pp.1486-1496, 2021.

[10] Belokurov, V., Kravtsov, A., "From Dawn Till Disc: Milky Way's Turbulent Youth Revealed By The APOGEE+Gaia Data", Monthly Notices of the Royal Astronomical Society, vol. 514, pp. 689-714, 2022.

[11] Sujatha, S., Babu, G. S. D., "Study of Open Cluster NGC 2509", Bulletin of the Astronomical Society of India, vol. 31, pp. 9-18, 2003.

[12] Carraro, G., Costa, E., "Photometry of the Five Marginally Studied Open Clusters Collinder 74, Berkeley 27, Haffner 8, NGC 2509, and VdB-Hagen 4", Astronomy and Astrophysics, vol. 464, pp. 573-580, 2007.

[13] Kharchenko, N.V., Piskunov, A.E., Roeser, S., Schilbach, E., Scholz, R.-D. "Global Survey of Star Clusters in the Milky Way, II. The Catalogue of Basic Parameters", Astronomy and Astrophysics, vol. 558, article id.A53, pp. 1-8, 2013.

[14] Joshi, Y. C., Dambis, A. K., Pandey, A. K., Joshi, S., "Study of Open Clusters within 1.8 kpc and Understanding the Galactic Structure", Astronomy and Astrophysics, vol. 593, article id.A116, pp. 1-13, 2016.

[15] Liu, L., Pang, X., "A Catalog of Newly Identified Star Clusters in Gaia DR2", The Astrophysical Journal Supplement Series, vol. 245, article id.A32, pp. 1-13, 2019.

[16] Siegel, M. H., LaPorte, S. J., Porterfield, B. L., Hagen, L. M. Z., Gronwall, C. A, "The Swift UVOT Stars Survey. III. Photometry and Color-Magnitude Diagrams of 103 Galactic Open Clusters", The Astronomical Journal, vol. 158, article id.A35, pp. 1-27, 2019.

[17] Cantat-Gaudin, T., Anders, F. "Clusters and Mirages: Cataloguing Stellar Aggregates in the Milky Way", Astronomy and Astrophysics, vol. 633, article id.A99, pp. 1-22, 2020.

[18] Cantat-Gaudin, T., Anders, F., Castro-Ginard, A., Jordi, C., Romero-Gómez, M., Soubiran, C., Casamiquela, L., Tarricq, Y., Moitinho, A., Vallenari, A., Bragaglia, A., Krone-Martins, A., Kounkel, M.., "Painting a Portrait of the Galactic Disc with its Stellar Clusters", Astronomy and Astrophysics, vol. 640, article id.A1, pp. 1-17, 2020.

[19] de Juan Ovelar, M., Gossage, S., Kamann, S., Bastian, N., Usher, C., Cabrera-Ziri, I., Dotter, A., Conroy, C., Lardo, C., "Extended Main Sequence Turnoffs in Open Clusters as Seen by Gaia - II. The Enigma of NGC 2509", Monthly Notices of the Royal Astronomical Society, vol. 491, pp. 2129-2136, 2020.

[20] Dias, W. S., Monteiro, H., Moitinho, A., Lépine, J. R. D., Carraro, G., Paunzen, E., Alessi, B., Villela, L. "Updated Parameters of 1743 Open Clusters Based on Gaia DR2", Monthly Notices of the Royal Astronomical Society, vol. 504, pp. 356-371, 2021.

[21] Gaia Collaboration, Vallenari, A., Brown, A. G. A., Prusti, T., de Bruijne, J. H. J., Arenou, F., Babusiaux, C., Biermann, M., Creevey, O. L., Ducourant, C., Evans, D. W., Eyer, L., Guerra, R., Hutton, A., Jordi, C., Klioner, S. A., Lammers, U. L., Lindegren, L., Luri, X., Mignard, F., Panem, C., Pourbaix, D., Randich, S., Sartoretti, P., Soubiran, C., Tanga, P., Walton, N. A., Bailer-Jones, C. A. L., Bastian, U., Drimmel, R., Jansen, F., Katz, D., Lattanzi, M. G., van Leeuwen, F., Bakker, J., Cacciari, C., Castañeda, J., De Angeli, F., Fabricius, C., Fouesneau, M., Frémat, Y., Galluccio, L., Guerrier, A., Heiter, U., Masana, E., Messineo, R., Mowlavi, N., Nicolas, C., Nienartowicz, K., Pailler, F., Panuzzo, P., Riclet, F., Roux, W., Seabroke, G. M., Sordo, R., Thévenin, F., Gracia-Abril, G., Portell, J., Teyssier, D., Altmann, M., Andrae, R., Audard, M., Bellas-Velidis, I., Benson, K., Berthier, J., Blomme, R., Burgess, P. W., Busonero, D., Busso, G., Cánovas, H., Carry, B., Cellino, A., Cheek, N., Clementini, G., Damerdji, Y., Davidson, M., de Teodoro, P., Nuñez Campos, M., Delchambre, L., Dell'Oro, A., Esquej, P., Fernández-Hernández, J., Fraile, E., Garabato, D., García-Lario, P., Gosset, E., Haigron, R., Halbwachs, J. -L., Hambly, N. C., Harrison, D. L., Hernández, J., Hestroffer, D., Hodgkin, S. T., Holl, B., Janßen, K., Jevardat de Fombelle, G., Jordan, S., Krone-Martins, A., Lanzafame, A. C., Löffler, W., Marchal, O., Marrese, P. M., Moitinho, A., Muinonen, K., Osborne, P., Pancino, E., Pauwels, T., Recio-Blanco, A., Reylé, C., Riello, M., Rimoldini, L., Roegiers, T., Rybizki, J., Sarro, L. M., Siopis, C., Smith, M., Sozzetti, A., Utrilla, E., van Leeuwen, M., Abbas, U., Ábrahám, P., Abreu Aramburu, A., Aerts, C., Aguado, J. J., Ajaj, M., Aldea-Montero, F., Altavilla, G., Álvarez, M. A., Alves, J., Anders, F., Anderson, R. I., Anglada Varela, E., Antoja, T., Baines, D., Baker, S. G., Balaguer-Núñez, L., Balbinot, E., Balog, Z., Barache, C., Barbato, D., Barros, M., Barstow, M. A., Bartolomé, S., Bassilana, J. -L., Bauchet, N., Becciani, U., Bellazzini, M., Berihuete, A., Bernet, M., Bertone, S., Bianchi, L., Binnenfeld, A., Blanco-

Cuaresma, S., Blazere, A., Boch, T., Bombrun, A., Bossini, D., Bouquillon, S., Bragaglia, A., Bramante, L., Breedt, E., Bressan, A., Brouillet, N., Brugaletta, E., Bucciarelli, B., Burlacu, A., Butkevich, A. G., Buzzi, R., Caffau, E., Cancelliere, R., Cantat-Gaudin, T., Carballo, R., Carlucci, T., Carnerero, M. I., Carrasco, J. M., Casamiquela, L., Castellani, M., Castro-Ginard, A., Chaoul, L., Charlot, P., Chemin, L., Chiaramida, V., Chiavassa, A., Chornay, N., Comoretto, G., Contursi, G., Cooper, W. J., Cornez, T., Cowell, S., Crifo, F., Cropper, M., Crosta, M., Crowley, C., Dafonte, C., Dapergolas, A., David, M., David, P., de Laverny, P., De Luise, F., De March, R., De Ridder, J., de Souza, R., de Torres, A., del Peloso, E. F., del Pozo, E., Delbo, M., Delgado, A., Delisle, J. -B., Demouchy, C., Dharmawardena, T. E., Di Matteo, P., Diakite, S., Diener, C., Distefano, E., Dolding, C., Edvardsson, B., Enke, H., Fabre, C., Fabrizio, M., Faigler, S., Fedorets, G., Fernique, P., Fienga, A., Figueras, F., Fournier, Y., Fouron, C., Fragkoudi, F., Gai, M., Garcia-Gutierrez, A., Garcia-Reinaldos, M., García-Torres, M., Garofalo, A., Gavel, A., Gavras, P., Gerlach, E., Geyer, R., Giacobbe, P., Gilmore, G., Girona, S., Giuffrida, G., Gomel, R., Gomez, A., González-Núñez, J., González-Santamaría, I., González-Vidal, J. J., Granvik, M., Guillout, P., Guiraud, J., Gutiérrez-Sánchez, R., Guy, L. P., Hatzidimitriou, D., Hauser, M., Haywood, M., Helmer, A., Helmi, A., Sarmiento, M. H., Hidalgo, S. L., Hilger, T., Hładczuk, N., Hobbs, D., Holland, G., Huckle, H. E., Jardine, K., Jasniewicz, G., Jean-Antoine Piccolo, A., Jiménez-Arranz, Ó., Jorissen, A., Juaristi Campillo, J., Julbe, F., Karbevska, L., Kervella, P., Khanna, S., Kontizas, M., Kordopatis, G., Korn, A. J., Kóspál, Á., Kostrzewa-Rutkowska, Z., Kruszyńska, K., Kun, M., Laizeau, P., Lambert, S., Lanza, A. F., Lasne, Y., Le Campion, J. -F., Lebreton, Y., Lebzelter, T., Leccia, S., Leclerc, N., Lecoeur-Taibi, I., Liao, S., Licata, E. L., Lindstrøm, H. E. P., Lister, T. A., Livanou, E., Lobel, A., Lorca, A., Loup, C., Madrero Pardo, P., Magdaleno Romeo, A., Managau, S., Mann, R. G., Manteiga, M., Marchant, J. M., Marconi, M., Marcos, J., Marcos Santos, M. M. S., Marín Pina, D., Marinoni, S., Marocco, F., Marshall, D. J., Martin Polo, L., Martín-Fleitas, J. M., Marton, G., Mary, N., Masip, A., Massari, D., Mastrobuono-Battisti, A., Mazeh, T., McMillan, P. J., Messina, S., Michalik, D., Millar, N. R., Mints, A., Molina, D., Molinaro, R., Molnár, L., Monari, G., Monguió, M., Montegriffo, P., Montero, A., Mor, R., Mora, A., Morbidelli, R., Morel, T., Morris, D., Muraveva, T., Murphy, C. P., Musella, I., Nagy, Z., Noval, L., Ocaña, F., Ogden, A., Ordenovic, C., Osinde, J. O., Pagani, C., Pagano, I., Palaversa, L., Palicio, P. A., Pallas-Quintela, L., Panahi, A., Payne-Wardenaar, S., Peñalosa Esteller, X., Penttilä, A., Pichon, B., Piersimoni, A. M., Pineau, F. -X., Plachy, E., Plum, G., Poggio, E., Prša, A., Pulone, L., Racero, E., Ragaini, S., Rainer, M., Raiteri, C. M., Rambaux, N., Ramos, P., Ramos-Lerate, M., Re Fiorentin, P., Regibo, S., Richards, P. J., Rios Diaz, C., Ripepi, V., Riva, A., Rix, H. -W., Rixon, G., Robichon, N., Robin, A. C., Robin, C., Roelens, M., Rogues, H. R. O., Rohrbasser, L., Romero-Gómez, M., Rowell, N., Royer, F., Ruz Mieres, D., Rybicki, K. A., Sadowski, G., Sáez Núñez, A., Sagristà Sellés, A., Sahlmann, J., Salguero, E., Samaras, N., Sanchez Gimenez, V., Sanna, N., Santoveña, R., Sarasso, M., Schultheis, M., Sciacca, E., Segol, M., Segovia, J. C., Ségransan, D., Semeux, D., Shahaf, S., Siddiqui, H. I., Siebert, A., Siltala, L., Silvelo, A., Slezak, E., Slezak, I., Smart, R. L., Snaith, O. N., Solano, E., Solitro, F., Souami, D., Souchay, J.,

Spagna, A., Spina, L., Spoto, F., Steele, I. A., Steidelmüller, H., Stephenson, C. A., Süveges, M., Surdej, J., Szabados, L., Szegedi-Elek, E., Taris, F., Taylor, M. B., Teixeira, R., Tolomei, L., Tonello, N., Torra, F., Torra, J., Torralba Elipe, G., Trabucchi, M., Tsounis, A. T., Turon, C., Ulla, A., Unger, N., Vaillant, M. V., van Dillen, E., van Reeven, W., Vanel, O., Vecchiato, A., Viala, Y., Vicente, D., Voutsinas, S., Weiler, M., Wevers, T., Wyrzykowski, Ł., Yoldas, A., Yvard, P., Zhao, H., Zorec, J., Zucker, S., Zwitter, T.,"Gaia Data Release 3: Summary of the Content and Survey Properties", Astronomy and Astrophysics, vol. 674, article id.A1, pp.1-22, 2023.

[22] Monteiro, H., Dias, W. S., "Distances and Ages From Isochrone Fits of 150 Open Clusters Using Gaia DR2 Data", Monthly Notices of the Royal Astronomical Society, vol. 487, pp. 2385-2406, 2019.

[23] King, I. R., "The Structure of Star Clusters. I. An Empirical Density Law", Astronomical Journal, vol. 67, pp. 471-485, 1962.

[24] Dias, W. S., Monteiro, H., Lépine, J. R. D., Prates, R., Gneiding, C. D., Sacchi, M., "Astrometric and Photometric Study of Dias 4, Dias 6, and Other Five Open Clusters Using Ground-Based and Gaia DR2 Data", Monthly Notices of the Royal Astronomical Society, vol. 481, pp. 3887-3901, 2018.

[25] Banks, T., Yontan, T., Bilir, S., Canbay, R., "Vilnius photometry and Gaia astrometry of Melotte 105", Journal of Astrophysics and Astronomy, vol. 41, article id.A6, pp. 1-24, 2020.

[26] Koç, S., Yontan, T., Bilir, S., Canbay, R., Ak, T., Banks, T., Ak, S., Paunzen, E., "A Photometric and Astrometric Study of the Open Clusters NGC 1664 and NGC 6939", The Astronomical Journal, vol. 163, article id.A191, pp. 1-22, 2022.

[27] Yontan, T., "An Investigation of Open Clusters Berkeley 68 and Stock 20 Using CCD UBV and Gaia DR3 Data", Astronomical Journal, vol. 165, article id.A79, pp. 1-20, 2023.

[28] Krone-Martins A., Moitinho A., "UPMASK: Unsupervised Photometric Membership Assignment in Stellar Clusters", Astronomy and Astrophysics, vol. 561, article id.A57, pp. 1-12, 2004.

[29] Bressan, A., Marigo, P., Girardi, L., Salasnich, B., Dal Cero, C., Rubele, S., Nanni, A., "PARSEC: Stellar Tracks and Isochrones with the Padova and Trieste Stellar Evolution Code", Monthly Notices of the Royal Astronomical Society, vol. 427, pp. 127-145, 2012.

[30] Sun, M., Jiang, B., Yuan, H., Li, J., "The Ultraviolet Extinction Map and Dust Properties at High Galactic Latitude", The Astrophysical Journal Supplement Series, vol. 254, article id.38, pp.1-12, 2021.

[31] Taşdemir, S, Yontan, T., "Analysis of the Young Open Cluster Trumpler 2 Using Gaia DR3 Data", Physics and Astronomy Reports, vol. 1, no. 1, pp. 1-10, 2023.

[32] Koç, S., Yontan, T., "Astrophysical Parameters of the Open Cluster Berkeley 6", Bitlis Eren Üniversitesi Fen Bilimleri Dergisi, vol. 12, no. 2, pp. 369-375, 2023.